\documentclass[12pt]{article} \usepackage{a4wide,amssymb}

\makeatletter

\@addtoreset{equation}{section} \makeatother

\def\be{\begin{equation}}
\def\ee{\end{equation}}
\def\bea{\begin{eqnarray}}
\def\eea{\end{eqnarray}}

 \def\Ov{{\mit\Omega}_{\rm v}}                     \def\Ev{E_{\rm v}}
 \def\lv{\ell_{\rm v}}     \def\Rv{R_{\rm  v}}     \def\nv{n_{\rm v}}
 \def\rhv{\rho_{\rm v}}    \def\rhc{\rho_{\rm c}}  
 \def\rhN{\rho_{\scriptscriptstyle \rm N}}
 \def\Tx{T_{\rm x}}        \def\TN{T_{\scriptscriptstyle \rm N}} 
 \def\mP{m_{\scriptscriptstyle \rm P}}
 \def\aa{g^*}              \def\aax{g^*_{\rm x}}
\begin{document}
\title{\bf Cosmic D-Strings and Vortons in Supergravity}
\author{Ph. Brax$^1$\footnote{brax@spht.saclay.cea.fr},
C. van de Bruck$^2$\footnote{C.vandeBruck@sheffield.ac.uk},
A. C. Davis$^3$\footnote{A.C.Davis@damtp.cam.ac.uk} \
 and Stephen C. Davis$^4$\footnote{sdavis@lorentz.leidenuniv.nl} \\
\\
\small \em ${}^1$ Service de Physique Th\'eorique, CEA/DSM/SPhT,
Unit\'e de recherche associ\'ee au CNRS, \\ \small \em
CEA--Saclay F--91191 Gif/Yvette cedex, France.
\\
\small \em ${}^2$ Department of Applied Mathematics, University of Sheffield,\\
\small \em Hounsfield Road, Sheffield S3 7RH, UK.
\\
\small \em ${}^3$ DAMTP, Centre for Mathematical Sciences, University of
Cambridge, \\ \small \em  Wilberforce Road, Cambridge, CB3 0WA, UK.
\\
\small \em ${}^4$ Instituut-Lorentz for Theoretical Physics, Postbus 9506,\\
\small \em NL--2300 RA Leiden, The Netherlands.
}

\maketitle
\begin{abstract}
Recent developments in string inspired models of inflation suggest
that D-strings are formed at the end of inflation. Within the
supergravity model of D-strings there are $2(n-1)$ chiral fermion
zero modes for a D-string of winding $n$. Using the bounds on the
relic vorton density, we show that D-strings with winding number
$n>1$ are more strongly constrained than cosmic strings arising in
cosmological phase transitions. The D-string tension of such
vortons, if they survive until the present, has to satisfy $8\pi
G_N \mu \lesssim p \, 10^{-26}$ where $p$ is the intercommutation
probability. Similarly, D-strings coupled with spectator fermions
carry currents and also need to respect the above bound. D-strings
with $n=1$ do not carry currents and evade the bound. We discuss
the coupling of D-strings to supersymmetry breaking. When a
single $U(1)$ gauge group is present, we show that there is an
incompatibility between spontaneous supersymmetry breaking and
cosmic D-strings. We propose an alternative mechanism for
supersymmetry breaking, which includes an additional $U(1)$, and
might alleviate the problem. We conjecture what effect this would
have on the fermion zero modes.
\end{abstract}

\section{Introduction}

The formation of cosmic strings appears to be a generic feature of
recent models of brane inflation arising from fundamental string
theory~\cite{tye,copeland}. Indeed, lower dimensional branes are
formed when a brane and anti-brane annihilate with the production
of $D3$ and $D1$ branes, or D-strings, favoured~\cite{mahbub}. It
has been argued that D-strings have much in common with cosmic
strings in supergravity theories and that they could be identified
with D-term strings~\cite{dvali}. Whilst cosmic strings in global
supersymmetric theories have been analysed 
before~\cite{susyCS,susybreak}, the study of such strings with local SUSY
is at an early stage since the presence of supergravity produces
added complications. These have recently been addressed in
refs.~\cite{dvali,binetruy,BBDD}. In ref.~\cite{BBDD} an
exhaustive study of fermion zero modes was performed. It was found
that due to the presence of the gravitino, the number of zero mode
solutions in supergravity is reduced in some cases. In particular,
it was found that there are no zero modes for F-term strings. For
BPS D-strings with winding number $n$ the number of chiral zero
modes is  $2(n-1)$, rather than $2n$ in the global case. For
winding number $1$ strings this was observed in
ref.~\cite{rachel}. When there are spectator fermions present, as
might be expected if the underlying theory gives rise to the
standard model of particle physics, then there are $n$ chiral zero
modes per spectator fermion.

Ordinary cosmic strings are either long strings or
loops~\cite{vilenkin&shellard}. When two strings meet they
intercommute with unit probability. When a long string
self-intersects it forms a long string and a loop. String loops decay
via gravitational radiation. This results in the string network rapidly
reaching a scaling solution, and the string loops never dominate the energy
density of the universe. This picture is changed in the presence of
fermion zero modes. The zero modes move along the string resulting in
the string carrying a current~\cite{witten}. Loops can be stabilised
by the presence of the current-carriers, giving rise to
vortons~\cite{vortons}. These can be used to constrain the
underlying theory~\cite{BCDT,CD}. For cosmic strings or
D-strings arising from fundamental string theory the intercommutation
probability is less than unity~\cite{Jackson}. This results in a
slower, denser string network with a higher string loop
density~\cite{mairi,Avgoustidis}. If such strings are current-carrying
then the resulting vorton density could be higher. We use bounds from both
nucleosynthesis, and the closure of the universe, to constrain
the D-string tension for vortons surviving until either Big Bang
Nucleosynthesis, or the present time. Such bounds are extremely stringent. Of
course the $n=1$ winding D-strings evade these bounds, unless
they couple to spectator fermions. Finally we discuss the
embedding of the D-strings in a broken SUSY environment. This is of
great phenomenological importance. We find obstacles to embedding
D-strings in broken SUSY with a single $U(1)$ gauge group. We discuss
a way round this by introducing an extra $U(1)$, and conjecture what
effect this will have on the fermion zero modes.

The letter is arranged as follows. In section 2 we describe the
properties of D-strings in supergravity and count the number of
zero modes. In section 3, we analyse the constraints on the
D-string tension coming from the relic density of vortons during
BBN and today. Finally in section 4 we present some results on the coupling
between D-strings and SUSY breaking.

\section{D-strings and their Fermion Zero Modes}

In this section we use the supergravity description of
D-strings by D-term strings~\cite{dvali}. Consider a supergravity
theory with fields $\phi^\pm$, charged under an Abelian gauge
group. The D-term bosonic potential
\begin{equation}
V = \frac{g^2}{2}(|\phi^+|^2 - |\phi^-|^2 - \xi)^2
\end{equation}
includes a non-trivial Fayet-Iliopoulos term $\xi$. Such a term is
compatible with supergravity provided the superpotential has
charge $-\xi$. Here the superpotential vanishes.  The minimum of
the potential is $|\phi^+|^2=\xi$. It is consistent to take the
cosmic string solution to be
\begin{equation} \phi^+ =
f(r) e^{in\theta}
\end{equation}
where $n$ is the winding number, and $\phi^-= 0$ since this minimises
the corresponding vector mass~\cite{ADPU}. The function $f(r)$
interpolates between $0$ and $\sqrt \xi$. The presence of a cosmic
string bends space-time. Its gravitational effects
lead to a deficit angle in the far away metric of
spacetime. In the following we consider the metric
\begin{equation}
ds^2=e^{2 B}(-dt^2 + dz^2) + dr^2 + C^2 d\theta^2
\end{equation}
for a cosmic string configuration. This is the most general
cylindrically symmetric metric, as discussed by Thorne~\cite{Thorne}.
Far away from the string the energy momentum tensor is approximately zero and
therefore $B$ is zero. Now we also find that
\begin{equation}
C =  C_1 r + C_0 + O(r^{-1}) \ ,
\end{equation}
When $C_1\ne 0$, the solution is a cosmic string solution with a
deficit angle $\Delta= 2\pi (1-C_1)$.

In supersymmetric theories, a cosmic string generally breaks all
supersymmetries in its core. BPS objects are an exception
to this rule, as they leave 1/2 of the original supersymmetry
unbroken. D-strings, in which we will be interested in this paper,
are an example of this. These strings have vanishing $T_{rr}$,
and the conformal factor $B$ is identically zero. Moreover for
D-strings one finds
\begin{equation}
\Delta = 2 \pi |n|\xi \ .
\end{equation}
 The supersymmetry algebra in four dimensions allows for 1/2
BPS configurations which saturate a BPS bound giving an equality
between the mass, i.e.\ the tension, and a central charge. Other
cosmic strings have higher tension and are not BPS, i.e.\ they
break all supersymmetries. This implies that $C_1$ is less than
the BPS case giving
\begin{equation}
\Delta\ge 2\pi\vert n \vert \xi \ .
\end{equation}

Let us now consider the fermion partners of the D-string bosonic
fields. These are the higgsino $\chi^+$, the gaugino $\lambda$,
and gravitino $\psi_{\mu}$. On D-strings, the number of zero mode
bound states is drastically affected by supergravity effects. We will discuss
D-term strings and then add spectator fields as might be the case when
embedding D-strings in a particle physics model.

Let us first consider the model in global
supersymmetry~\cite{susyCS}. In this case the zero
modes can be constructed explicitly. The zero mode solutions have
definite chirality, in the sense that they are eigenstates of
$\sigma^3$.
We find that for large $r$, there is one solution for which
$\lambda$ and $\chi$ decay exponentially. For positive
chirality zero modes, we find that this solution has the form
\begin{equation}
\lambda \sim r^{m} e^{i(m+1/2)\theta} \ , \ \ \chi^+ \sim r^{n-1-m}
e^{i(n-m-1/2)\theta} \ , \label{mm}
\end{equation}
near the origin.
For the solution to be normalisable using the ${\cal L}^2$ norm, the
integer $m$ must satisfy $0 \leq m \leq n-1$.   Denoting the number of
normalisable zero mode solutions of chirality $\pm$ by $N^\pm$, we find
\begin{equation}
N^+=2n \ , \ \ N^-= 0 \label{DstrN}
\end{equation}
for $n>0$. For negative $n$ the two chiralities are interchanged,
and there are $2|n|$ zero modes with negative chirality.

In contrast to the global case, we must now include the gravitino
field. In supergravity, zero modes are obtained once a gauge choice
has been made for the supersymmetry transformations. It is
particularly convenient to work in the $\bar \sigma^\mu \psi_\mu=0$
gauge. In this gauge, the gravitino has three degrees of freedom, each
being a Weyl fermion. The equations of motion for the gravitino in a
BPS background imply that $\psi_t=\psi_z=0$. The remaining degree of
freedom is a single Weyl fermion,
$\Psi = \sigma^r \bar \psi_r - \sigma^\theta \bar \psi_\theta$. It has
the same norm as the other non-gravitino Weyl fermions. Moreover the
field $\Psi$ has the same chirality as the ordinary fermions.

Near $r=0$ the general solution to the fermion field
equations, for positive chirality states, has the leading order behaviour
\begin{equation}
\vert \chi^+\vert  = c_1 r^{n-m-1} \ , \vert \lambda \vert = c_2 r^{m}
\ , \vert \Psi\vert = c_3 r^{m-1} \ ,
\end{equation}
where the $c_i$ are constants. At infinity just one combination of
solutions decays exponentially, and (for $m\geq 0$) is the only
normalisable solution there. In general, the form of this
combination of solutions near $r=0$ will be given by the above
expression with all $c_i \neq 0$. Hence if the solution is to be
normalisable everywhere, we must have $m \leq n -1$, $m \geq
0$ and $m \geq 1$ (the last condition comes from the gravitino field
$\Psi$). Without the gravitino we would need $n-1 \geq m \geq 0$.
Including it we lose the $m=0$ mode as it cannot be normalisable
close to the origin. The rest of the zero mode tower is preserved.
Notice that this effect is purely local and does not depend on the
behaviour of the fields at infinity. For $n>0$ the number of zero modes is
therefore
\begin{equation}
N^+=2(n-1) \ , \ N^-=0 \ .
\end{equation}
This can be confirmed using the generalised index theorem~\cite{BBDD,index}.

The D-string model can be augmented with new spectator fields.
These fields may appear in particle physics models.  We add to the
D-strings new fields $\Phi_i$ and a superpotential
\begin{equation}
W= \sum_{i=1}^M \frac{a_i}{2} \phi^+ \Phi_i^2
\end{equation}
These fields are massive outside the string core. In the
string background, we have $\Phi_i=0$. The fermion $\chi_i$ associated with
$\Phi_i$ does not then couple to the string sector, and so it can be
analysed independently.  Applying the index theorem (with $n>0$) we find
\begin{equation}
N^+=0 \ , \ N^-=nM \ .
\end{equation}
The presence of these chiral zero modes has been obtained in
ref.~\cite{binetruy}. The above result also holds for global SUSY, and
is not affected by the inclusion of SUGRA effects. The
result will be unaffected by supersymmetry breaking.

\section{Cosmic Vortons}

The existence of fermion zero modes in the string core can have
dramatic consequences for the properties of cosmic strings. For
example, the zero modes can be excited and will move along the
string, resulting in the string carrying a current~\cite{witten}.
The direction of the charge carriers is determined by the
chirality of the zero modes. If they all have the same chirality,
they will all move in the same direction. The current will be
maximal, and chiral. An initially weak current on a string loop is
amplified as the loop contracts and could become sufficiently
strong to prevent the string loop from decaying. In this case a
stable loop, or vorton~\cite{vortons}, is produced. Vortons are
classically stable~\cite{Carter}, so if they do decay, then its
probably via quantum mechanical tunnelling, resulting in them
being very long lived. In the case of fermion zero modes it has
been shown that they are indeed very stable~\cite{currstab},
particularly when the zero modes are chiral such as those under
consideration here~\cite{Blanco-Pillado}. The density of vortons
is tightly constrained by cosmology. In particular, if vortons are
sufficiently stable so that they survive until the present time
then the universe must not be vorton dominated. However, if
vortons only survive a few minutes then they can still have
cosmological implications since the universe must be radiation
dominated at nucleosynthesis. These requirements have been used to
constrain such models~\cite{BCDT,CD} . In the case of strings
arising from a fundamental string theory there is another
consideration to take into account. In ref.~\cite{Jackson} it was
shown that for such strings the probability of intercommutation,
$p$, is less than unity. For D-strings it was estimated that $p$
is between $0.1<p<1$. Strings with lower $p$ are slower and have a
denser string network, resulting in the number of loops being
proportional to $p^{-1}$~\cite{mairi,Avgoustidis}. Consequently we
need to recalculate the vorton bounds taking into account the
reduced intercommutation probability. A string loop has two
conserved quantum numbers; $N$ is the topologically conserved
integral of the phase current and $Z$ is the particle number,
which are identically equal for chiral currents. This results in
the angular momentum quantum number $J=N^2$. The energy of the
vorton is $\Ev\simeq \lv \mu$, where $\mu$ is the string tension.
Taking the vorton to be approximately circular, the radius is
$\Rv=\lv/2\pi$, which gives $\lv\simeq(2\pi)^{1/2}N\mu^{-1/2}$.
Thus we obtain an estimate of the vorton mass energy as \be
\Ev\simeq N\mu^{1/2}\ , \label {energy} \ee where we are assuming
the classical description of the string dynamics. This is valid
only if the length $\lv$ is large compared with the relevant
quantum wavelengths, which is satisfied if $N$ is sufficiently
large. A loop that does not satisfy this requirement will never
stabilise as a vorton. We can now calculate the vorton abundance, 
following~\cite{CD} but remembering that the intercommutation
probability is less than unity.
Assuming that the string becomes current carrying at a scale $\Tx$
then one expects that thermal fluctuations will give rise to a
current. For strings arising from a cosmological phase transition,
$\Tx$ is the transition temperature. Whilst for strings arising in
brane inflation models $\Tx$ is the relevant string
scale~\cite{cline}. In both cases we expect $\Tx \simeq
\mu^{1/2}$. Loops with $N \gg 1$ should satisfy the minimum length
requirement, otherwise the loop is doomed to lose all its energy
and disappear. The total number density of small loops with length
of order $L_{\rm min}$, the minimum length for vortons, will be
not much less than the number density of all closed loops and
hence $n\simeq L_{\rm min}^{-3}(T) p^{-1}$. The typical length
scale of string loops at the transition temperature, $L_{\rm
min}(\Tx)$, is considerably greater than the relevant thermal
correlation length, ${\Tx}^{-1}$, resulting in a modified string
loop evolution. Loops present at the time of the current
condensation then satisfy $L\geq L_{\rm min}(\Tx)$, and reasonably
large values of the quantum number $N$ build up. If $\lambda$ is
the wavelength of the fluctuation of the carrier field then $N
\simeq L/\lambda$, where $\lambda\simeq\Tx^{-1}$. Thus, one
obtains \be N \simeq L_{\rm min}(\Tx)\Tx \gg 1 \ . \ee For
current condensation during the early friction-dominated regime of
string network evolution this requirement is always satisfied.
Therefore, the vorton mass density is $\rhv\simeq {N \mu^{1/2}
\nv}$. In the friction-dominated regime the string is interacting
with the surrounding plasma. We can estimate $L_{\rm min}$ in this
regime as the typical length scale below which the microstructure
is smoothed, with the dominant mechanism being Aharonov-Bohm
scattering. This length scale is $(\mP
\mu)^{1/2}/[(\aa)^{1/4}T^{5/2}]$, where $\aa(T)$ is the effective
number of massless degrees of freedom for the plasma. This then
gives the quantum number, $N\simeq \mP^{1/2}/(\aax \mu)^{1/4}$,
giving the number density of mature vortons $\nv\simeq \aa T^3
\mu^{3/4}/[(\aax)^{1/4}\mP^{3/2} p]$. The resulting mass
density of the relic vorton population is \be \rhv\simeq
\left(\frac{\mu}{\sqrt{\aax} \, \mP}\right) \frac{\aa T^3}{p} \ ,
\label{plus 31} \ee We are now in a position to bound the energy
scale of formation of current carrying strings.

The standard cosmological model requires
the universe to be radiation dominated at the time of
nucleosynthesis. Thus the energy density in vortons at that time,
$\rhv(\TN)$, should have been small compared with the background
energy density in radiation, $\rhN\simeq \aa \TN^4$. Assuming that carrier
condensation occurs during the friction damping regime, this gives
\be \left(\frac{\mu}{p \sqrt{\aax} \, \mP}\right) \ll \TN \ , \ee
where $\aax$ is the effective number of
degrees of freedom at the time of current condensation. If we make the rather
conservative assumption that vortons only survive for a few minutes,
which is all that is needed to reach the nucleosynthesis epoch, we
obtain a fairly strong restriction on the theory:
\be 8\pi G_N \mu \lesssim p \sqrt{\aax} \, 10^{-22}  \ee
where $G_N$ is Newton's constant. Taking $\aax \simeq 10^2$
in the early universe we obtain
\be 8\pi G_N \mu \lesssim p \, 10^{-21}  \ . \ee

If vortons are sufficiently stable to survive until the present
epoch one can find a more stringent constraint by requiring that
the vortons do not dominate the energy density today. We impose
that $\Ov \equiv  \rhv/\rhc \leq 1$, where $\rhc$ is the closure
density. Inserting our earlier estimate for the vorton density, we
can derive the dark matter constraint. This gives 
\be 8\pi G_N \mu\lesssim p \sqrt{\aax} \, 10^{-27} 
\ \simeq p \, 10^{-26} \ . \ee
This result is based on the assumptions that the vortons  are stable
enough to survive until the present day.
However, we expect it to be realistic for
the chiral case since such vortons are classically and quantum
mechanically very stable. These constraints put strong limits on
D-strings with fermion zero modes. We note that $n=1$ D-strings
would evade these constraints, but higher winding number strings
and D-strings with spectator fermions would be subject to the
above limits. These limits are much stronger than estimates of the
string tension for D-strings formed in brane inflation models.

\section{Coupling D-strings to Spontaneously Broken Supersymmetry}

In the previous section we have studied the properties of a vorton
producing string network. In a realistic context, one must take into account
that supersymmetry will be broken; coupling a SUSY-breaking sector
to a D-string is non-trivial. The SUSY-breaking will change the microphysics
of the string, and we also expect it to alter the conductivity of the
strings and the corresponding vorton constraints.

In order for the Fayet-Iliopoulos term $\xi$ to be present in the
theory, the superpotential must have charge $-\xi$. It can
therefore be written as
\begin{equation}
W= (\phi^+)^{-\xi} W_{SB}(\ldots)
\end{equation}
where the function $W_{SB}$ depends on only uncharged combinations of
fields, and so is uncharged itself. This means that in a string
background the gravitino mass $m_{3/2} \propto W \propto e^{-i n\xi \theta}$.
For most string solutions this will not be single valued (there is no
problem before SUSY-breaking since $W=0$ then).

In general, when supersymmetry is broken, there will be a
discontinuity in the phase of $m_{3/2}$ around the strings. Domain walls will
form along this discontinuity. A string network will therefore
develop a network of domain walls between the different strings. The
walls will try to contract, pulling the strings together until they
either annihilate, or form higher winding number strings with single
valued $m_{3/2}$ around them. Alternatively the walls will not
contract quickly enough and we will be left with a network of domain
walls, which will conflict with observation.

For the above model the gravitino mass will only be single valued
if $n\xi$ is an integer, resulting in a supermassive
string~\cite{vilenkin&shellard}. The string deficit angle is then
$\Delta =2 \pi (1-C_1) \geq 2\pi |n|\xi$. This implies, for $n
\neq 0$, that the deficit angle is greater or equal to $2\pi$. If
$\Delta = 2\pi$ the space outside the string is topologically
equivalent to $S^1 \times \mathcal{R}^2$, which is unphysical. On
the other hand if $\Delta > 2\pi$ there is a curvature singularity
at finite distance from the string where $C(r)=0$, which is again
unphysical. This seems to suggest that D-strings and SUSY-breaking
cannot be combined in the same model.

A possible way round this problem would be to include an
additional $U(1)$ in the theory, with a corresponding
Fayet-Iliopoulos term $\tilde \xi$. We also include an additional
field $\rho$ which has unit charge under the new gauge group. For
the superpotential to have the right gauge dependence it must now
have the form
\begin{equation}
W= (\phi^+)^{-\xi} \rho^{-\tilde \xi} W_{SB}(\ldots) \ .
\end{equation}
We assume that after SUSY-breaking $\rho$ gains a VEV, and we
suppose that it winds around the string  
$\rho \propto e^{im\theta}$. This implies 
$m_{3/2} \propto e^{-i (n\xi + m\tilde \xi)\theta}$.  
Hence if $n$, $m$ and $n\xi + m\tilde \xi$ are all integers, the
string will be single valued. If $\xi /\tilde \xi$ is rational it is
always possible to find suitable $n$ and $m$. It is energetically 
favourable for the winding $m$ to settle down to a value for which all
fields, and $m_{3/2}$, are single-valued (assuming such a value
exists). Otherwise there will be a sharp jump in the phase of $\rho$
at some value of $\theta$. The energy of the resulting domain wall
means that such a configuration is disfavoured. For some values of 
$\xi /\tilde \xi$, string solutions with $n=1$ will not have
single-valued $m_{3/2}$ for any choice of $m$. In this case walls will
inevitably form between different strings. The domain wall tension will then
pull the strings together until a configuration with a higher $n$,
which does allow single-valued $m_{3/2}$, is produced. Alternatively
the force between the strings may be too weak, and we will instead be
left with a network of domain walls.

If $W_{SB}$ and the coupling of the extra D-term are small, we expect their
contribution to the string mass per unit length, and also its
deficit angle to be small. Thus, assuming a self consistent model
of the above form can be constructed, D-strings are compatible
with SUSY-breaking. Hence the vorton constraints from section 3
will still apply. The above situation is similar to an $SO(10)$
string at electroweak symmetry breaking~\cite{so10}. There it
seems that the electroweak Higgs will not be single valued around
the string. However, by taking into account the effect of a $U(1)$
subgroup of the electroweak sector, the problem is avoided.

Since the form of the string is changed by SUSY-breaking, it
is possible that its conductivity, and the corresponding vorton
bounds will also be altered. We see that SUSY-breaking couples
more fermion fields to the string, and that the solutions also have
more Yukawa couplings which wind around the string. From the index
theorem~\cite{BBDD,index} we find that both these factors tend to increase
string conductivity. This means that the strings could become
conducting after SUSY-breaking, even if they do not conduct
before. The corresponding vorton bounds in this case are weaker, as
discussed in refs.~\cite{BCDT,CD}. On the other hand we expect that
there will be some cases where the conductivity is reduced by SUSY-breaking.

Vorton constraints arising from spectator zero modes will not be affected by
the above arguments. The spectator sector fermions do not couple to any
of the string field fermions, so string conductivities in the two
sectors are independent. Of course the spectator fermions could couple
to other fields involved in SUSY-breaking, and this could alter the
number of spectator zero modes.

\section{Conclusions}

In this letter we have seen that winding number $n$ cosmic D-strings
have  $2(n-1)$ chiral fermion zero modes in their core. Similarly cosmic
D-strings with $M$ spectator fermions  have $nM$ chiral zero modes in
addition to those arising from the actual D-string itself. These
zero modes can be excited and give rise to currents, which can
stabilise string loops. We have investigated the resulting
bounds on stable string loops, or vortons, taking into account the
fact that if the string arises from a higher dimensional underlying
theory, the intercommutation probability will be reduced. This leads
to more stringent bounds than their field theory counterparts. For vortons
which are absolutely stable we found that
$8\pi G_N \mu \lesssim p \, 10^{-26}$ where $p$ is the intercommutation
probability. For vortons that survive only a few
minutes, probably decaying via quantum mechanical tunnelling, we found 
$8\pi G_N \mu \lesssim p \, 10^{-21}$. However, as discussed previously
vortons formed from cosmic D-strings are likely to be particularly 
stable, so the dark matter constraint is probably the most realistic in
this case. Both bounds are much stronger than bounds on the string
tension arising from structure formation, and the estimates from brane
inflation models. The latter is model dependent, but estimates suggest
that the D-string tension is between 
$10^{-12} \lesssim G_N \mu \lesssim 10^{-6}$~\cite{tye}.

We have also investigated the effect of supersymmetry breaking  on
cosmic D-strings. For BPS strings there is an incompatibility
between soft supersymmetry breaking, as might arise from gluino
condensation, and cosmic D-strings. This arises because the
gravitino is charged under the Fayet-Iliopoulos term resulting in
superpotentials having charge $-\xi$. The requirement that the
gravitino mass is single valued implies that $\xi$ must be
quantised, giving a deficit angle bigger than $2\pi$, which is
unphysical. This seems to be an underlying problem with cosmic
D-strings. To avoid this we have proposed  another way of breaking
supersymmetry in D-string theories, which includes an additional
$U(1)$ symmetry. In order to ascertain the effect of this on the
fermion zero modes requires a detailed model, which is beyond the
scope of this paper and will be left for future investigation.
However, it is unlikely such a mechanism would destroy the zero
modes and could even result in introducing extra zero modes at the
scale of supersymmetry breaking. If this were to happen the vorton
bounds would change to take into account the two scales along the
lines of ref.~\cite{CD}.

\subsection*{Acknowledgements}
ACD is grateful for CEA Saclay for hospitality during some of this work.
This work was supported in part by PPARC, the Netherlands Organisation
for Scientific Research (NWO), the RTN European programme MRTN-CT-2004-503369
and the ESF Coslab programme.


\begin{thebibliography}{99}
\bibitem{tye}
  S.~Sarangi and S.~H.~H.~Tye,
  Phys.\ Lett.\ B {\bf 536} (2002) 185  [hep-th/0204074]; \\
  N.~Jones, H.~Stoica and S.~H.~H.~Tye,
  JHEP {\bf 0207} (2002) 051  [hep-th/0203163].
\bibitem{copeland}
  E.~J.~Copeland, R.~C.~Myers and J.~Polchinski,
  JHEP {\bf 0406} (2004) 013  [hep-th/0312067].
\bibitem{mahbub}
  M.~Majumdar and A.~C.~Davis,
  JHEP {\bf 0203} (2002) 056
  [hep-th/0202148].
\bibitem{dvali}
  G.~Dvali, R.~Kallosh and A.~Van Proeyen,
  JHEP {\bf 0401} (2004) 035  [hep-th/0312005].
\bibitem{susyCS}
  S.~C.~Davis, A.~C.~Davis and M.~Trodden,
  Phys.\ Lett.\ B {\bf 405} (1997) 257  [hep-ph/9702360].
\bibitem{susybreak}
  S.~C.~Davis, A.~C.~Davis and M.~Trodden,
  Phys.\ Rev.\ D {\bf 57} (1998) 5184  [hep-ph/9711313].
\bibitem{binetruy}
  P.~Binetruy, G.~Dvali, R.~Kallosh and A.~Van Proeyen,
  Class.\ Quant.\ Grav.\  {\bf 21} (2004) 3137 [hep-th/0402046].
\bibitem{BBDD}
  Ph.~Brax, C.~van de Bruck, A.~C.~Davis and S.~C.~Davis,
  JHEP (to appear) [hep-th/0604198].
\bibitem{rachel}
  R.~Jeannerot and M.~Postma,
  JHEP {\bf 0412} (2004) 043 [hep-ph/0411260].
\bibitem{vilenkin&shellard}
  A.~Vilenkin and E.~P.~S.~Shellard, `Cosmic Strings and Other
  Topological Defects' Cambridge University Press (1994); \\
  M.~B.~Hindmarsh and T.~W.~B.~Kibble,
  Cosmic strings,
  Rept.\ Prog.\ Phys.\  {\bf 58} (1995) 477
  [hep-ph/9411342].
\bibitem{witten}
  E.~Witten,
  Nucl.\ Phys.\ B {\bf 249} (1985) 557.
\bibitem{vortons}
  R.~L.~Davis and E.~P.~S.~Shellard,
  Phys.\ Lett.\ B {\bf 209} (1988) 485.
\bibitem{BCDT}
  R.~H.~Brandenberger, B.~Carter, A.~C.~Davis and M.~Trodden,
  Phys.\ Rev.\ D {\bf 54} (1996) 6059  [hep-ph/9605382].
\bibitem{CD}
  B.~Carter and A.~C.~Davis,
  Phys.\ Rev.\ D {\bf 61} (2000) 123501  [hep-ph/9910560].
\bibitem{Jackson}
  M.~G.~Jackson, N.~T.~Jones and J.~Polchinski,
  JHEP {\bf 0510} (2005) 013
  [hep-th/0405229].
\bibitem{mairi}
  M.~Sakellariadou,
  JCAP {\bf 0504} (2005) 003 [hep-th/0410234].
\bibitem{Avgoustidis}
  A.~Avgoustidis and E.~P.~S.~Shellard,
  Phys.\ Rev.\ D {\bf 71} (2005) 123513
  [hep-ph/0410349].
\bibitem{ADPU}
  A.~Achucarro, A.~C.~Davis, M.~Pickles and J.~Urrestilla,
  Phys.\ Rev.\ D {\bf 66} (2002) 105013
  [hep-th/0109097].
\bibitem{Thorne}
  K.~S.~Thorne,
  Phys.\ Rev.\ {\bf 138} (1965) B251; \\
  R.~Gregory,
  Phys.\ Rev.\ Lett.\  {\bf 59} (1987) 740.
\bibitem{index}
  S.~C.~Davis, A.~C.~Davis and W.~B.~Perkins,
  Phys.\ Lett.\ B {\bf 408} (1997) 81 [hep-ph/9705464].
\bibitem{Carter}
  B.~Carter and X.~Martin,
  Annals Phys.\  {\bf 227} (1993) 151
  [hep-th/0306111].
\bibitem{currstab}
  S.~C.~Davis, W.~B.~Perkins and A.~C.~Davis,
  Phys.\ Rev.\ D {\bf 62} (2000) 043503
  [hep-ph/9912356].
\bibitem{Blanco-Pillado}
  J.~J.~Blanco-Pillado, K.~D.~Olum and A.~Vilenkin,
  Phys.\ Rev.\ D {\bf 66} (2002) 023506
  [hep-ph/0202116].
\bibitem{cline}
  N.~Barnaby, A.~Berndsen, J.~M.~Cline and H.~Stoica,
  JHEP {\bf 0506} (2005) 075
  [hep-th/0412095].
\bibitem{so10}
  A.~C.~Davis and S.~C.~Davis,
  Phys.\ Rev.\ D {\bf 55} (1997) 1879
  [hep-ph/9608206]; \\
  W.~B.~Perkins and A.~C.~Davis,
  Nucl.\ Phys.\ B {\bf 406} (1993) 377.

\end{thebibliography}
\end{document}